\documentclass[conference]{IEEEtran}
\IEEEoverridecommandlockouts
\usepackage{cite}
\usepackage{amsmath,amssymb,amsfonts}
\usepackage{eqparbox}
\usepackage{arydshln}
\usepackage{algorithmic}
\usepackage{graphicx}
\usepackage{textcomp}
\usepackage{xcolor}
\usepackage{multirow}
\usepackage{tablefootnote}

\def\BibTeX{{\rm B\kern-.05em{\sc i\kern-.025em b}\kern-.08em
    T\kern-.1667em\lower.7ex\hbox{E}\kern-.125emX}}

\newcommand{\linebreakand}{%
  \end{@IEEEauthorhalign}
  \hfill\mbox{}\par
  \mbox{}\hfill\begin{@IEEEauthorhalign}
}   
    
\begin{document}

\title{YA-DA: YAng-Based DAta Model for Fine-Grained IIoT Air Quality Monitoring\\
}

\author{
\IEEEauthorblockN{
Yagmur Yigit\IEEEauthorrefmark{1},
Khayal Huseynov\IEEEauthorrefmark{1}, 
Hamed Ahmadi\IEEEauthorrefmark{2}, and 
Berk Canberk\IEEEauthorrefmark{3} \IEEEauthorrefmark{4}}
\IEEEauthorblockA{\IEEEauthorrefmark{1} Department of Computer Engineering, Istanbul Technical University, Turkey \\
\IEEEauthorrefmark{2} Department of Electronic Engineering, University of York, United Kingdom \\
\IEEEauthorrefmark{3} Department of Artificial Intelligence and Data Engineering, Istanbul Technical University, Turkey \\
\IEEEauthorrefmark{4} School of Computing, Engineering and The Build Environment, Edinburgh Napier University, United Kingdom \\
Email: \{yigity20, canberk\}@itu.edu.tr, khayal.huseynov@btsgrp.com,\\hamed.ahmadi@york.ac.uk, B.Canberk@napier.ac.uk}
}

\maketitle

\begin{abstract}

With the development of industrialization, air pollution is also steadily on the rise since both industrial and daily activities generate a massive amount of air pollution. Since decreasing air pollution is critical for citizens' health and well-being, air pollution monitoring is becoming an essential topic. Industrial Internet of Things (IIoT) research focuses on this crucial area. Several attempts already exist for air pollution monitoring. However, none of them are improving the performance of IoT data collection at the desired level. Inspired by the genuine Yet Another Next Generation (YANG) data model, we propose a YAng-based DAta model (YA-DA) to improve the performance of IIoT data collection. Moreover, by taking advantage of digital twin (DT) technology, we propose a DT-enabled fine-grained IIoT air quality monitoring system using YA-DA. As a result, DT synchronization becomes fine-grained. In turn, we improve the performance of IIoT data collection resulting in lower round-trip time (RTT), higher DT synchronization, and lower DT latency.

\end{abstract}

\begin{IEEEkeywords}
Industrial Internet of Things (IIoT), Air Quality Monitoring, Yet Another Next Generation (YANG) Data Model, Digital Twins (DT). 
\end{IEEEkeywords}

\section{Introduction}
\label{sec:intro}

The fast urbanization of the human population and the increasing industrial activities have a consequential impact on global air quality. The rapid development of cities significantly increases the pollutants produced by the ever-increasing number of vehicles, traffic volume, industrial sites, etc. According to data from the World Health Organization (WHO), ninety-nine percent of the global population is exposed to air pollution at different levels \cite{WHO}. Air quality significantly impacts daily life, and prolonged exposure to air pollution can injure the respiratory, immune, and nervous systems and even result in the development of cancer \cite{TransactionsAirQ}. Therefore, air quality monitoring is becoming essential to protect and improve the life quality of citizens in overcoming these problems. Thanks to the Internet of Things (IoT) technological advancement, the number of IoT-connected devices is rapidly increasing. According to the forecasts, the number of IoT-connected devices worldwide will reach around 29.4 billion by 2030 \cite{Gartner}. One of the most significant applications of the IoT is the Industrial Internet of Things (IIoT). Gartner Report also envisions that IIoT devices will populate more than half of IoT devices soon. IIoT focuses on the connectivity of intelligent machines in different domains such as transportation, manufacturing, healthcare systems, etc. Therefore, it is an essential concept for air quality monitoring.

\begin{figure*}[t]
    \centering
    \includegraphics[width=6.5in]{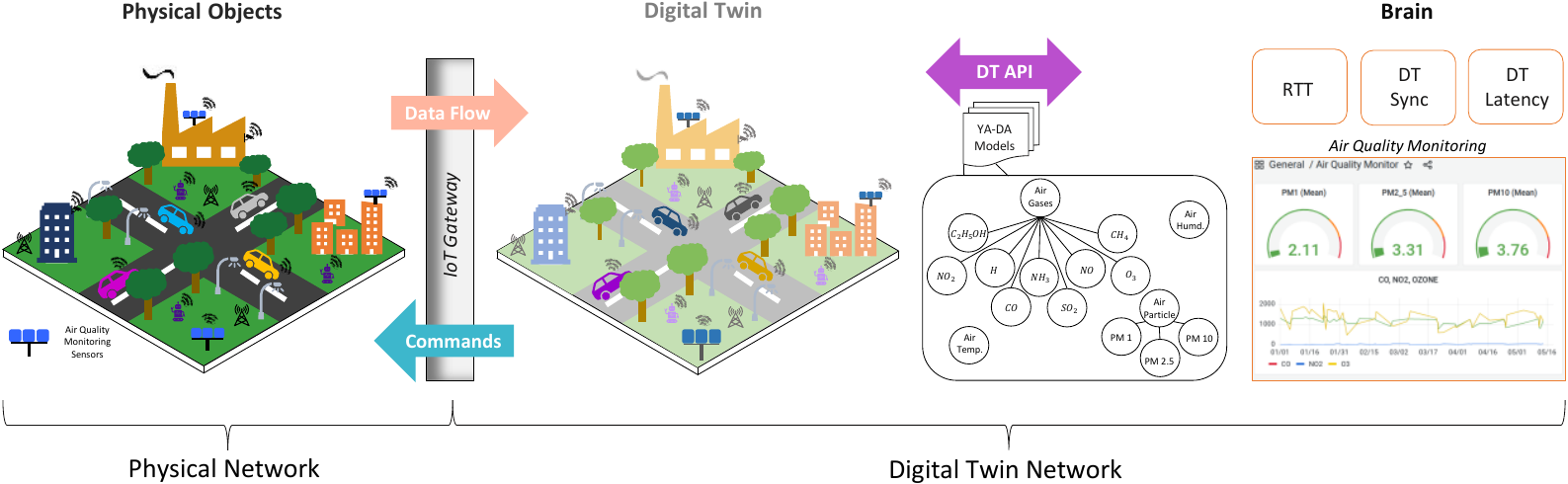}
    \caption{The proposed system architecture.}
    \label{fig:system}
\end{figure*}

Combining the robust data model-driven automation needed in IIoT with device and network efficiency is crucial. IIoT devices have interfaces to manipulate connected equipment. To correctly manage resources on each device, all relevant data model interfaces must be known during development. Yet Another Next Generation (YANG) is a data modeling language for device configuration. It is rich enough to express the characteristics of the related data model. The YANG model is a popular topic in the networking industry as it is easy to learn and can describe complex programmatic interfaces \cite{RFC6020}. It provides many different mechanisms for IoT devices \cite{IoTYang}. For instance, it supports distributed authorship to use modules from various sources together; language extensions, and model extensions without centralized control; it provides lifecycle, conformance model, platform-specific model deviations for actual deployment, etc. 
Furthermore, interoperability is essential between IoT devices and the rest of the IP world. In next-generation network systems, all devices will connect, called the Internet of Everything; network elements and IoT devices will configure together. At this point, YANG can be used as the data modeling language to manage network elements, and services \cite{Yang8199IoT}. Another promising technology is the digital twin (DT), which has rapidly become an essential part of industrial applications. It bridges the gap between physical and digital spaces by creating a real-time virtual replica of the objects in the system. DT provides a high-performance simulation model for analysis, testing, configuration, and optimization. Recently, DT-based visualization has become an important issue, especially with the smart city concept. With the help of DT-based visualization, air pollution levels can be visualized in real time across the city.

The amount of data being processed is increasing rapidly in parallel with the advancement of technology. For this reason, only relevant data should be processed in the related system so that the system can use it more efficiently for specific applications such as air quality monitoring, anomaly detection, etc. \cite{AutoML}. Thus, the overload in the system is reduced. Data collection efficiency is essential for IIoT systems since it can seriously affect the freshness of the data captured and the decision-making process. Besides, efficient IIoT data collection reduces energy consumption, latency, network lifetime, and overall cost. However, existing air pollution monitoring works have high Round-trip time (RTT), relatively low DT synchronization, and relatively high DT delay. Therefore, IoT-based air quality monitoring was inefficient. In this paper, we are improving the performance of IoT data collection by proposing a new data model called the YAng-based DAta model (YA-DA), based on the YANG model. By taking advantage of DT technology, we propose a DT-enabled fine-grained IIoT air quality monitoring system using YA-DA. After implementing the YA-DA data structure, our approach has lower RTT, higher DT synchronization, and lower DT latency. We will provide more detail in the following sections.

The rest of the paper is organized as follows. Section~\ref{sec:relWork} provides a survey of related studies in the literature. The proposed solution and the performance evaluation are explained in Section~\ref{sec:proposed} and Section~\ref{sec:perfEval} respectively. We conclude the paper in Section~\ref{sec:conc}.

\section{Related Work}
\label{sec:relWork}

Since data collection is an essential part of sensor devices and related technologies such as IIoTs and wireless sensor networks (WSN), there are many works published in this research area to propose different models, architectures, and contributions in the domains. I. Ali \textit{et al.} gave a comprehensive review of the data collection in the IoT, WSN, and sensor cloud research domains \cite{collectionSurvey}. The findings show that studies on data collection are relatively consistent with stable output in the last five years. They also provided critical research challenges and future research directions for data collection in related research domains. Z. Sheng \textit{et al.} proposed a lightweight approach to enable device management of wireless sensor devices by taking advantage of the recent development of IPv6-based open standards for accessing wireless resource-constrained networks \cite{2015IoTJ}. They developed a prototype to test the performance of the proposed approach in managing wireless sensor devices, thus contributing to IoT data collection development. W. Osamy \textit{et al.} suggested an intelligent data collection technique to determine energy-aware disjoint dominating sets that work as data collection nodes in each round to improve overall WSN lifetime \cite{2021SensorJ}. They mainly focused on energy conservation to enable the efficient functioning and lifetime of WSNs. They provided the efficiency of the proposed technique mathematically and in simulation. N. Koroniotis \textit{et al.} presented a smart airport cyber twins security-oriented IIoT testbed SAir-IIoT \cite{SAir}. It comprises multiple heterogeneous IIoT devices and communication protocols that can be accessed remotely as a service, automatically interconnected with each other. They proposed a data management tool that dynamically collects, analyzes, and labels network-acquired and telemetry IIoT data. In another work, S. Chen \textit{et al.} proposed a privacy-preserving data collection and computation offloading scheme for efficient and secure big sensory data collection in fog-assisted IoT \cite{2020Transac}. The simulation results show that the proposed method is an efficient data collection and computation offloading scheme with a strong privacy preservation property. C. T. Cheng \textit{et al.} designed the concurrent data collection trees for IoT applications \cite{2017Tree}. Their results show that the proposed tree structures perform better in data collection processes.

Some works focus on air quality monitoring taking advantage of DT technology in the literature.
G. Mylonas \textit{et al.} gave a detailed review of current DT research in the field of intelligent cities and also drew parallels with the applications of Industry 4.0 \cite{2021SurvetDTAir}. They emphasized the importance of pollution monitoring in the smart city domain and reviewed the current works of pollution monitoring. They also presented the open challenges of city-scale DTs. G. Schrotter \textit{et al.} presented a city-scale DT of Zurich to facilitate use in several applications like air and noise pollution monitoring \cite{Schrotter2020TheDT}. They presented a study that promotes the active digital participation of citizens in urban planning procedures by visualizing a variety of data covering various applications, such as thermal monitoring of the city. They note the significance of using open spatial data in encouraging dissemination and developing new applications. H. Lehner \textit{et al.} discussed a new strategy for producing the three-dimensional city model and the geographical data of Vienna, which completely rethinks geographical data in the geographical data workflows \cite{Lehner2020DigitalGV}. After discussing the effect of the level of detail of three-dimensional city representation, they provided the simulation results that calculate the pollution dispersion inside a city. In another work, T. Nochta \textit{et al.} provided a city-scale DT prototype presented to tackle congestion, air pollution, growth management, and the limited capacity of the local energy infrastructure issues in the Cambridge city region \cite{2021CityScale}. They underlined the importance of city-scale DT to reflect the specifics of the urban and socio-political context. Y. Liu \textit{et al.} proposed a federated learning-based aerial-ground air quality sensing framework for fine-grained three-dimensional air quality monitoring and forecasting \cite{UAV21}. They also proposed a graph convolutional neural network-based extended short-term memory model for ground sensing systems. Overall, most current air pollution monitoring DT works focus on the smart city domain, particularly the city-scale DT.

None of the previous studies specifically covered IIoT-based air quality monitoring in terms of data collection. So, their performance was relatively low. Thanks to the proposed approach, the performance in the particular DT-based air pollution monitoring model will be increased. The improved performance metrics with our solution are DT synchronization, DT latency, and RTT.

\section{Proposed Solution}
\label{sec:proposed}

Our proposed system consists of a physical network with physical objects and a digital twin network with digital twins and brain components. The proposed architecture is shown in Fig.~\ref{fig:system}. A replica of physical objects and synchronization between physical objects and digital twins are built. The IoT gateway collects sensor data and provides cyber-physical interaction between the physical objects and the digital twin. Then, sensor data is taken by the digital twin. After that, the air quality monitoring system takes the only desired data thanks to the software-defined YA-DA data model.

\begin{table}[t]
\caption{Semantic model of air monitoring sensors based on \\ the YA-DA data structure
\label{tab:table1}} 
\centering
\begin{tabular}{l}

\hline
\rule{0pt}{1.5em}\textbf{Software-defined YA-DA Data Model for Air Quality Monitoring} \\ [0.4em]
\hline\\
container AirParticleURI $\{$
\\
\hspace{0.1in} description ``Air Monitoring Particle Sensor''\\
\hspace{0.1in} list value $\{$ \\
\hspace{0.25in} key ``pm2.5-data'';\\
\hspace{0.25in} leaf pm1-data $\{$type air-sensor;$\}$ \\
\hspace{0.25in} leaf pm2.5-data $\{$type air-sensor;$\}$ \\
\hspace{0.25in} leaf pm10-data $\{$type air-sensor;$\}$ \\
\hspace{0.1in} $\}$ 
\\     
$\}$ 
\\ \\ 
\hdashline\\ 
container AirTemperatureURI $\{$
\\
\hspace{0.1in} description ``Air Monitoring Temperature Sensor''\\
\hspace{0.1in} leaf value $\{$type air-sensor;$\}$ \\
$\}$ 
\\ \\ 
\hdashline\\ 
container AirHumidityURI $\{$
\\
\hspace{0.1in} description ``Air Monitoring Humidity Sensor''\\
\hspace{0.1in} leaf value $\{$type air-sensor;$\}$ \\
$\}$ 
\\ \\ 
\hdashline\\ 
container AirGasesURI $\{$
\\
\hspace{0.1in} description ``Detectable Air Quality Gases''\\
\hspace{0.1in} container value $\{$ \\
\hspace{0.25in} leaf carbon-monoxide-data $\{$type air-sensor;$\}$ \\
\hspace{0.25in} leaf nitric-oxide $\{$type air-sensor;$\}$ \\
\hspace{0.25in} leaf nitrogen-dioxide $\{$type air-sensor;$\}$ \\
\hspace{0.25in} leaf sulphur-dioxide $\{$type air-sensor;$\}$ \\
\hspace{0.25in} leaf ethanol $\{$type air-sensor;$\}$ \\
\hspace{0.25in} leaf hydrogen $\{$type air-sensor;$\}$ \\
\hspace{0.25in} leaf ammonia $\{$type air-sensor;$\}$ \\
\hspace{0.25in} leaf methane $\{$type air-sensor;$\}$ \\
\hspace{0.25in} leaf ozone $\{$type air-sensor;$\}$ \\
\hspace{0.1in} $\}$ 
\\     
$\}$ 
\\ \\ 
\hline

\end{tabular}
\end{table}
\begin{table*}[t]
\caption{ The specifications of datasets
\label{tab:table2}} 
\centering
\begin{tabular}{|cl|c|c|c|}
\hline 
\multicolumn{2}{|c|}{Dataset}   & Number of Features & Number of Samples & Number of Used Samples \\  \hline
\multicolumn{1}{|c|}{\multirow{6}{*}{TON-IoT \cite{dataset1}}} & Fridge Activity       & 6                  & 59944             & 1000                   \\ \cline{2-5} 
\multicolumn{1}{|c|}{}                         & Garage Door Activity  & 6                  & 59587             & 800                   \\ \cline{2-5} 
\multicolumn{1}{|c|}{}                         & GPS Tracker Activity  & 6                  & 58960             & 2200                   \\ \cline{2-5} 
\multicolumn{1}{|c|}{}                         & Modbus Activity       & 7                  & 51106             & 2000                   \\ \cline{2-5} 
\multicolumn{1}{|c|}{}                         & Motion Light Activity & 6                  & 59488             & 1000                   \\ \cline{2-5} 
\multicolumn{1}{|c|}{}                         & Thermostat Activity   & 6                  & 52774             & 1000                   \\ \hline
\multicolumn{2}{|c|}{AQ\&U  \cite{dataset2}}                                            & 11                 & 522000            & 2000                   \\ \hline
\end{tabular}
\end{table*}

In the proposed system, we have spatial and temporal complexity. Spatial complexity is related to memory usage in the system. It needs to minimize by pulling only the relevant data into the air quality monitoring system. Temporal complexity pertains to end-to-end delay. It should be reduced to enhance the quality of service of the system. The software-defined YA-DA data model provides fine-grained air quality monitoring and reduces the computational time in the system by preventing the system from overloading. Thus, the system's operation becomes more efficient, and we reduce the spatial and temporal complexity.

\subsection{Software-defined YAng-based DAta Model (YA-DA)}

YANG is a data modeling language for configuration, and monitoring \cite{RFC7950}. It is originally designed to model data for the Network Configuration (NETCONF) protocol. NETCONF is a network management protocol. It provides basic programming features for convenient and robust automation of network services. After a YANG module has defined data hierarchies, it can use for NETCONF-based operations. These can include state data, configuration, notifications, and remote procedure calls. All data sent between a NETCONF client and a server can be described. NETCONF is just one of many protocols that can be used with YANG. For instance, RESTCONF and the Constrained Application Protocol (CoAP) Management Interface (CoMI), etc. NETCONF is more CPU and memory efficient than RESTful and gRPC services \cite{Muge}.

The hierarchical organization of data is modeled as a tree in which each node has a name-value pair or a set of child nodes in YANG. It provides clear and brief descriptions of nodes along with the interaction between these nodes. A YANG module includes a combination of related definitions. It can import definitions from external modules and contain explanations from submodules. Moreover, the hierarchy can extend by allowing a module to add data nodes to the order defined in another module.

By taking the above advantages of the YANG model, we designed the YA-DA data model to improve the performance of the IIoT data collection. We developed the YA-DA model in tree data structure because tree data structure performs better in IIoT data collection. We defined four main data node types in the YA-DA data model as in the YANG for data modeling: leaf nodes, leaf-list nodes, container nodes, and list nodes.

\begin{itemize}
\item \textit{Leaf Node} contains at most one instance in the data tree. A leaf has a value but no child nodes.
\item \textit{Leaf-List Node} defines a set of uniquely identifiable nodes rather than a single node. Each node has a value but no child nodes.
\item \textit{Container Node} is used to group related nodes in a subtree. A container has no value but rather a set of child nodes of any type, such as leaves, lists, containers, leaf lists, etc.
\item \textit{List Node} defines a sequence of list entries. Each entry is defined like a container and is uniquely identified by the values of the key leaves. A list can define multiple key leaves and may contain any number of child nodes of any type, including leaves, lists, containers, etc.
\end{itemize}
The semantic model of air quality monitoring sensors based on YA-DA data structure is shown in table~\ref{tab:table1}.

\begin{figure}[htbp]
    \centering
    \includegraphics[width=1.1in]{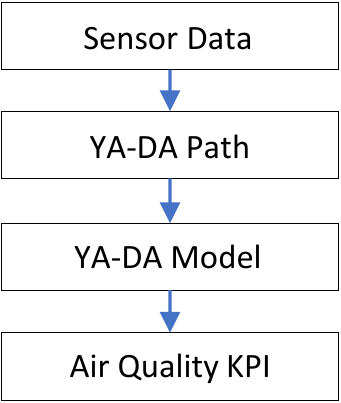}
    \caption{Hierarchy between YA-DA terms.}
    \label{fig:YANG}
\end{figure}

Since the amount of sensor data in the system is more than desired and needed, the air quality monitoring system can become very slow. It is necessary to import only the required data into the system. Therefore, we also use a software-defined YA-DA model to accomplish this. In this work, we defined an air quality KPI. Thanks to the YA-DA Paths, the system receives only the relevant data from the digital twin. Thus, we get an increasingly efficient, fine-grained system. The hierarchy between the mentioned terms can be better represented in Fig.~\ref{fig:YANG}.

\section{Performance Evaluation}
\label{sec:perfEval}

In this work, Microsoft Azure DT (ADT) is a platform we use as a service tool to provide twin graphs of the physical objects \cite{Azure}. ADT capabilities are DT Definition Language (DTDL), an open modeling language, live representation, input plugin, and output plugin. Using DTDL, we can define the digital models that represent our physical entities. The digital models can pinpoint semantic relationships between the entities. We define sensor elements using DTDL, and it has four main parts:

\begin{itemize}
\item \textit{Property} defines an entity's state. The property value is both writable and storable, meaning its historical value can be read at any time. 
\item \textit{Telemetry} shows measured values obtained through device sensor readings and is not stored in the ADT. Thus, it can stream through outsourced time-series applications for monitoring or time-based analysis.
\item \textit{Relationship} represents different semantic meanings between interfaces.
\item \textit{Component} describes something that is a part of the interface and does not need a separate entity in the twin graph.
\end{itemize}
The above explanation is not a full DTDL definition of a sensor entity but contains main parts. As a result, we have connected the twins into a graph that reflects their interactions. For data analysis and storage, the ADT platform is able to stream the data through an external output plugin \cite{Yagmur}. After implementing the twin models with predefined interfaces and relations among entities, we transmitted data to the brain part of the proposed system as the output plugin of ADT using the DT API.

We tested the performance of the software-defined YA-DA data model. To this end, we used TON-IoT and AQ\&U as two real-world datasets \cite{dataset1} \cite{dataset2}. The ToN-IoT dataset is comprised of heterogeneous data sources to evaluate the fidelity and efficiency of different applications for IoT and IIoT. The AQ\&U dataset is created to monitor air quality in Salt Lake City. The properties of these datasets are as in table~\ref{tab:table2}. We constituted a new dataset of ten thousand samples by taking particular samples from each dataset.

We reconfigured the sensors of the YA-DA data model according to the created dataset. Since we used an available dataset, not real-time data, we tested the proposed approach in the digital twin and brain areas of the digital twin network section of the proposed system. We set the DT synchronization to be between 0 and 1. The value of 1 means full synchronization, and 0 implies out-of-synchronization. We compared the proposed solution's performance implementation with and without the YA-DA data model. The results showed that all the required sensors to monitor air quality are classified successfully. As can be seen in table~\ref{tab:tablesync}, the system has better DT synchronization when using the YA-DA data model. Thus, DT synchronization becomes fine-grained.

\begin{table}[t]
\caption{The Comparison of DT Synchronization  
\label{tab:tablesync}}
\centering
\begin{minipage}{\linewidth} \centering
\begin{tabular}{|l|l|l|} 
\hline
\multirow{2}{*}{\begin{tabular}[c]{@{}l@{}}Number of\\Nodes\end{tabular}} & \multicolumn{2}{l|}{DT Synchronization $^{(*)}$}  \\ 
\cline{2-3}
                                                                          & \begin{tabular}[c]{@{}l@{}}with \\YA-DA\end{tabular} & \begin{tabular}[c]{@{}l@{}}without\\YA-DA\end{tabular}                                                                          \\ 
\hline
4                                                                         & 0.65                                                 & 0.47                                                                                                                            \\ 
\hline
6                                                                         & 0.61                                                 & 0.42                                                                                                                            \\ 
\hline
16                                                                        & 0.74                                                 & 0.57                                                                                                                            \\
\hline
\end{tabular}
\vspace{-10pt}
\footnotetext{\hspace{20pt}\scriptsize {$^{(*)}$The interval for the value is [0,1], in which closeness to 0 }}
\footnotetext{\hspace{20pt}\scriptsize{indicates poor and closeness to 1 represents a desirable outcome.}}
\end{minipage}

\end{table}

\begin{figure}[htbp]
    \centering
    \includegraphics[width=3.2in]{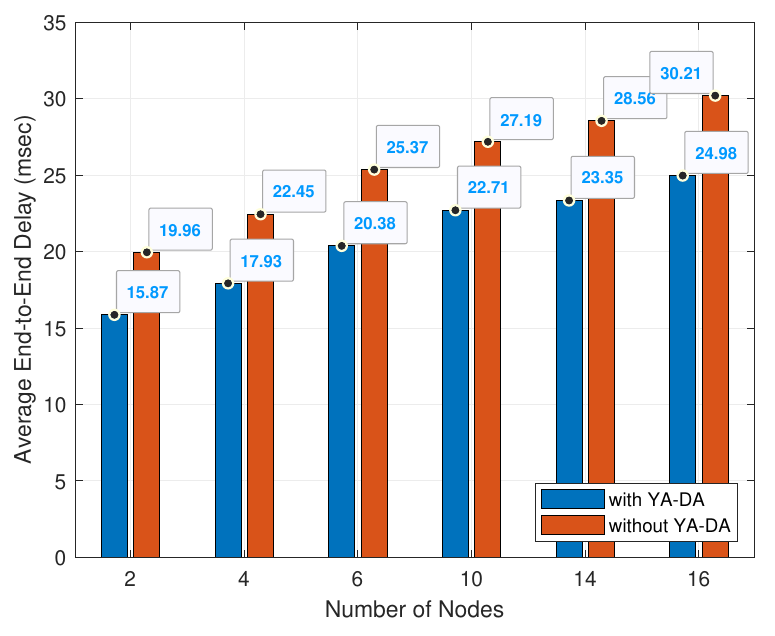}
    \caption{The comparison of average end-to-end delay.}
    \label{fig:e2e}
\end{figure}

Then, we tested the average end-to-end delay between the digital twin and brain areas of the digital twin network section of the proposed system. The results are shown in Fig.~\ref{fig:e2e}. Since the YA-DA data model has a tree structure, its average end-to-end delay is better. Hence, we can provide that the DT network has lower latency using YA-DA data model.

\begin{figure}[htbp]
    \centering
    \includegraphics[width=3.2in]{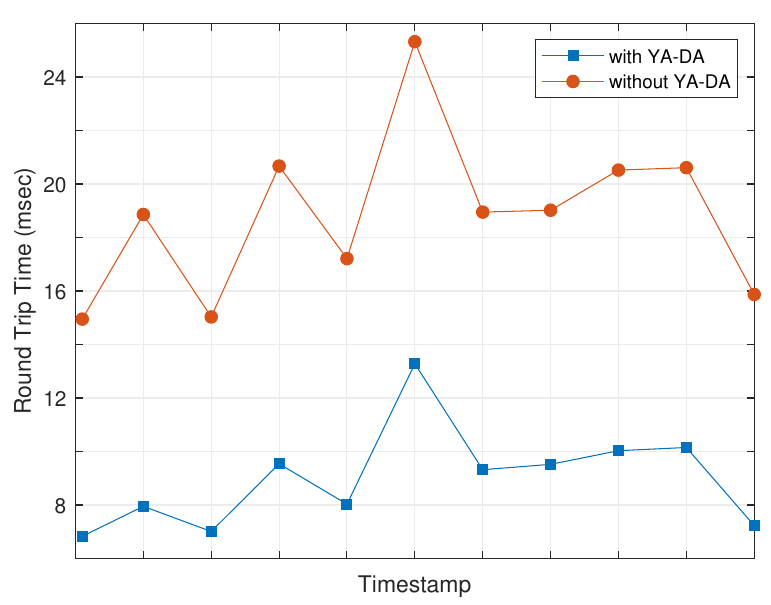}
    \caption{The comparison of RTT results.}
    \label{fig:RTT}
\end{figure}

After that, we examined the performance of the digital twin network. Fig.~\ref{fig:RTT} depicts the RTT result of the network. The results showed that the data payload became around forty percent better when we used the software-defined YA-DA data model based on YANG. 
The reason for this result is that we pulled data into the system using the YA-DA data model instead of all raw data. Moreover, the network load is reduced by getting specific data using YA-DA.

\begin{figure}[htbp]
    \centering
    \includegraphics[width=3.3in]{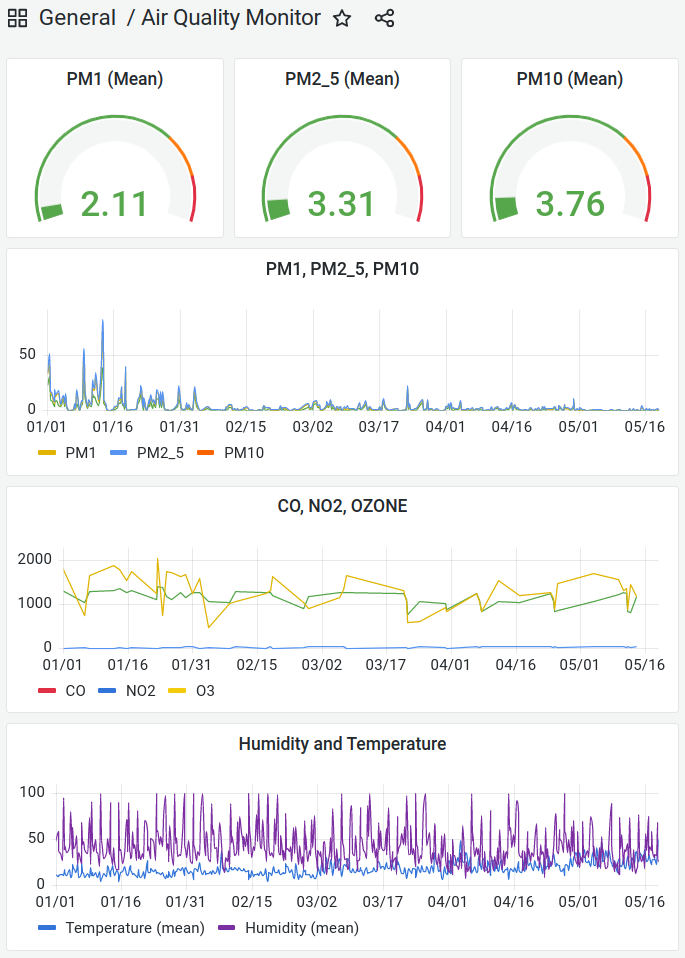}
    \caption{The monitoring of air quality data in the system.}
    \label{fig:monitor}
\end{figure}

Finally, we scrutinized the air quality data in the monitoring system of the brain part. We used Grafana to visualize and analyze our air quality data. Grafana is a web-based, open-source analytics and interactive visualization program \cite{Grafana}. Fig.~\ref{fig:monitor} demonstrates the graphical representation of the air quality data. This data includes particulate matter, air gases, humidity, and temperature values. According to the values, it can be stated that the air quality in this region is at an acceptable level.

\section{Conclusion}
\label{sec:conc}

In this paper, we design a YAng-based DAta model (YA-DA) to improve the performance of IIoT data collection inspired by the authentic YANG data model. DT synchronization becomes fine-grained thanks to the YA-DA data structure. Here, we propose a DT-enabled fine-grained IIoT air quality monitoring system by taking advantage of DT and IIoT technologies. Thus, we increase the efficiency of the air quality monitoring system. Our simulation results showed that our solution successfully reduces RTT and DT latency and improves DT synchronization thanks to the software-defined YA-DA data model.

\section*{Acknowledgment}
Yagmur Yigit would like to thank the DeepMind Scholarship Programme and the ITU-Turkcell Graduate Research Scholarship Programme for their support. This paper is supported by the Royal Society under Grant IES\textbackslash R3\textbackslash213169.

\bibliographystyle{IEEEtran}


\end{document}